# MPIFA: A Modified Protocol Independent Fairness Algorithm for Community Wireless Mesh Networks

C. H Widanapathirana, *Student Member IEEE*, Bok-Min Goi, Sim Moh Lim, *Member IEEE*

**Abstract** – Community Wireless Mesh Networks (WMN) is a paradigm in wireless communication of 21$^{st}$ centuary as means of providing high speed braodband access. Un-cooperative nodes, both selfish and malicious proves to be a significant threat in Community WMN that require a solution independent of routing protocols being used. We propose to implement Modified PIFA (MPIFA), an Improved version of Protocol Independent Fairness Algorithm (PIFA) proposed by Younghwan Yoo, Sanghyun and P. Agrawal [6] with ability to cater specific requirements in Community WMN. MPIFA has malicious nodes detection rate improvement of 50% when nodes demonstrate low probabilistic malicious behavior of 10% to circumvent the security measures in place. Improvements were also made to reduce false malicious node detections to 4% when node-to-node link failures occur in Community WMN.

## I. INTRODUCTION

Wireless Mesh Network (WMN) is an innovative network concept that has captured the attention of the scientific community which going through a paradigm shift from traditional centralized wireless systems such as cellular networks and wireless local area networks (LANs) to cooperative multi-hop wireless networks. In this work, the focus is on Community Wireless Mesh Networks which apply the WMN concept to connect independent users in a large geographical area. Fault tolerance, setup simplicity, robustness in WMN makes it an ideal candidate capable of providing high speed broadband network access to rural and urban communities with a cheaper cost and higher reliability [1].

Various social entities occupy the nodes in Community WMN and consequently, cooperative behavior among all the users to forward each other's packets cannot be assumed at all times. Existing routing protocols such as Ad hoc on-Demand Distance Vector (AODV) or Dynamic Source Routing (DSR) with the assumption of inherent node cooperation is challenged when implemented in Community WMN because cooperative behavior could not be assumed with social entities.

C.H Widanapathirana was with Faculty of Engineering, Multimedia University, Malaysia as an undergraduate student. He is now with Dialog Telekom, Sri Lanka (phone: +94772357496 e-mail: chathchw@gmail.com).

Dr Goi Bok Min is the associated dean and a senior lecturer of Faculty of Engineering, Multimedia University, Malaysia (phone: +60383125334 e-mail: bmgoi@mmu.edu.my).

Assoc Professor Sim Moh Lim was a senior lecturer in Faculty of Engineering, Multimedia University, Malaysia. He is now with Motorola Technology (M) Sdn Bhd, Bayan Lepas, 11900, Penang.(e-mail: mslim2003@gmail.com)

Two main types of uncooperative nodes could be observed in Community WMN, one type being selfish nodes and predominantly, malicious nodes which are economically rational nodes whose objective is to maximize its own welfare [2]. Malicious nodes introduce interruptive behavior to the network and try to avoid the fairness protocols and cheat the network. Practical implementations of outdoor Community WMN in large geographical areas often suffer with link failures in node-to-node communications. Malicious users in the community could demonstrate unpredictable behavior where they act probabilistically to drop packets. These specific challenges in Community WMN require a smarter algorithm to effectively identify malicious nodes.

In this paper we propose the Modified Protocol Independent Fairness Algorithm (MPIFA) to encourage user fairness while identifying and removing malicious nodes from the network. Initially we implement the Original PIFA algorithm to Community WMN to ensure its ability to perform in WMN and subsequently introduce improvements to cater the specific needs of Community WMN. Finally we analyze and compare the performance between Original PIFA and MPIFA using simulations.

## II. RELATED WORK

Node fairness research is fairly limited for WMN due to the novelty of the concept. ANTSEC, WATCHNT and ANTREP system put forth by Mogre et al. [3] discuss about node fairness in WMN but use an ANTSEC routing algorithm which makes it difficult to use with exciting WMN routing.

Packet forwarding fairness of nodes is discussed for another similar ad-hoc network type called mobile ad-hoc network (MANET) in few publications. There are two types of fairness algorithms being proposed for MANET. Credit based algorithms with centralized management and inherent fairness encouragement, was first proposed by Buttyan and Hubaux [4] in a system based on a virtual currency, called *nuglets*. Using *nuglets*, the authors proposed two payment models: the Packet Purse Model (PPM) and the Packet Trade Model (PTM). An alternative to PTM-PPM algorithm is SPIRIT (simple cheat proof credit based system) [2] which uses a credit management server called *CCS* with *receipts* from the nodes being transmitted to CCS for every message exchanged between nodes. This concept introduces a large additional overhead to the network. As opposed to that, Reputation based algorithms uses a localized management for detecting misbehaviors and takes punitive actions against the nodes such as in tools proposed by Marti et al. [5], Watchdog, which identifies misbehaving nodes, and Pathrater, which selects routes that avoid the identified nodes.

## III. PROTOCOL INDEPENDENT FAIRNESS ALGORITHM (PIFA)

In order to encourage nodes to cooperate and detect malicious nodes regardless of routing protocol being used in the community WMN, we suggest using the Protocol Independent Fairness Algorithm (PIFA) which was developed for MANET by Yoo et al. [6].

Even though there are significant differences between MANET and WMN in terms of routing methods, node resource constraints (fewer resources for mobile devices) as well as the user mobility, we hypothesize to use the basic algorithm introduced in PIFA to achieve the node fairness in WMN. Since both network technologies share the common ad-hoc characteristics and PIFA is fully independent of the routing protocols being used it is possible to implement PIFA in WMN even though there are significant differences.

PIFA is a credit based, centralized node fairness algorithm where nodes are encouraged to gather credits by help forwarding others' packets and use those credits to generate and forward their own packets. This concept encourages users to participate in the network operations as opposed to selfishly save resources. With the mechanism in the algorithm, the nodes which try to drop packets and cheat network management by reporting false packet forwarding information can be isolated and blacklisted from the network.

Implementation of PIFA is twofold where user nodes and a central management server (CMS) operate collectively.

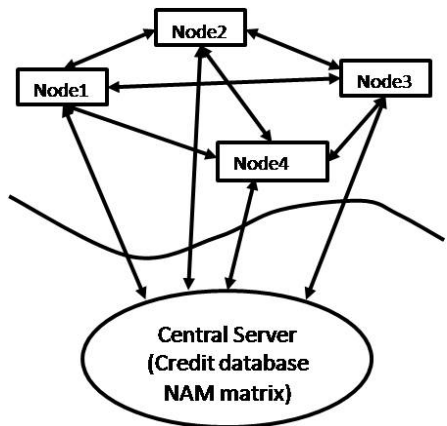

Fig 1: Basic structure of the algorithm

a.) User nodes

Each user node keeps a small database of data traffic (packets) that associated with other nodes. This traffic information will be transmitted to the CMS by nodes in regular intervals (CMS updating frequency) in a simple message which would look like the following.

| RID | NID | SEQ | I | O | S | T | OFN |

Fig 2: Report message to CMS

TABLE 1
CONTENT OF THE REPORT MESSAGE

| | |
|---|---|
| RID- | ID of a reporter |
| NID- | ID of a neighbor node |
| SEQ- | The sequence number of the current node's reports. |
| I- | Number of input packets from the neighbor |
| O- | Number of output packets to the neighbor |
| S- | Number of packets starting at the current node among output packets to the neighbor |
| T- | Number of packets terminated at the current node among input packets from the neighbor |
| OFN- | Number of packets originated from the neighbor itself among input packets from the neighbor |

Since there are multiple neighbors to a node, each node will issue messages pertaining to transactions with every neighboring node.

b.)   Central Management Server(CMS)

CMS is a server (or a powerful node) connected to the network where all the messages from nodes are stored, analyzed and management decisions are made. CMS contains two databases namely, credit database which contains credit information of each node and Number of Alleged Manipulations (NAM) database which contains node data manipulation allegations. When CMS receives all the messages from nodes in an update cycle, it will carry out 3 basic tests to check the validity of the messages being sent. If an unfair node drops packets while changing its counters to cover the packet drops maliciously these tests can detect these particular nodes.

First, the number of output packets from a node towards a particular neighbor node should be equal to the number of input packets received by neighbor node. Let $Q_{a,b}$ denote the **Q** field of a message whose **RID** and **NID** are $a$ and $b$, where nodes $a$ and $b$ are neighbors to each other. Then,

$$O_{a,b} = I_{b,a} \quad (1)$$

Second, if $F_a$ is the number of packets forwarded by node $a$ during a period and $b$ is the set of adjacent nodes of node $a$, then

$$F_a = \sum_b I_{a,b} - \sum_b T_{a,b} = \sum_b O_{a,b} - \sum_b S_{a,b} \quad (2)$$

The difference between the total number of input packets($\sum I$) and the total number of terminated packets ($\sum T$) at a node is equal to the number of forwarding packets as well as the difference between the total number of output packets ($\sum O$) and the total number of starting packets ($\sum S$).

$F_a$ is then used by the CMS to award credits for node $a$ by,

$$F_a \times \beta - \sum S_a \times \delta \quad (3)$$

amount, where $\beta$ and $\delta$ are constants determined by the network management to balance the credit so that nodes are rewarded appropriate amount of credit for forwarding packets and pay the appropriate amount when generating. If the credit of a node drops below a certain level, CMS will send warning messages to the node to encourage it to participate more in the packet forwarding process and if the total credit drops below lower threshold, the particular node will be banned from sending packets until it gathers enough credit by forwarding others packets. By using PIFA algorithm, nodes refrain from deliberately avoid packet forwarding because they need to earn credits to send their own packets. Initially, CMS assigns a fixed amount of credits to all nodes so that network could begin operations.

Malicious nodes could manipulate **F,** the number of forwarding packets by changing $\sum T$ and $\sum S$. Final test will compare number of packets originated from the neighbor itself among input packets from the neighbor, OFN of a certain node with number of starting packets, S of that neighbor sent to this node.

$$S_{a,b} = OFN_{b,a} \quad (4)$$

Test could detect any changes done by a malicious node to deliberately mislead CMS. This is done because commonly neighbor node does not have a strong motive to cheat on the network by supporting a particular malicious node.

**Malicious nodes detection**

Once a discrepancy is detected between two nodes messages (failed tests (1) (2) (3) (4) ), CMS could not unambiguously determine the malicious node (*a* or *b*). CMS assume both nodes are malicious and charge both these nodes for the time being and update the Number of Alleged Manipulations (NAM) data base for a single charge.

$$m_{a,b} = m_{a,b} + 1$$
$$m_{b,a} = m_{b,a} + 1 \quad (5)$$

This is called NAM increase policy (5) of the network. After each update cycle, CMS checks for nodes with NAM count higher than a threshold value and these nodes are blacklisted from the network determined as malicious. These nodes would not be allowed to join the network operation again.

|   | a | b | c | d | … | Total |
|---|---|---|---|---|---|---|
| a | … | $m_{a,b}$ | $m_{a,c}$ | $m_{a,d}$ | … | $\sum m_{a,i}$ |
| b | $m_{b,a}$ | … | $m_{b,c}$ | $m_{b,d}$ | … | $\sum m_{b,I}$ |
| c | $m_{c,a}$ | $m_{c,b}$ | … | $m_{c,d}$ | … | $\sum m_{c,I}$ |
| d | $m_{d,a}$ | $m_{d,b}$ | $m_{d,c}$ | … | … | $\sum m_{d,I}$ |

Fig 2: NAM database

In order to stop innocent nodes from getting penalized by NAM update policy, CMS in cooperates a NAM decrease policy (6). This is based on the assumption that a malicious node highly likely to repeat the attempts to deceive others: thus if nodes *a* and *b* report inconsistent information just now, other nodes' NAMs that have been accumulated in connection with nodes *a* and *b* before are reduced by half.

$$m_{i,a} = m_{i,a}/2 ,$$
$$m_{i,b} = m_{i,b}/2 , \text{ for all 'i' except (a,b)} \quad (6)$$

IV. MPIFA FOR PERFORMANCE OPTIMIZATION

Protocol independent fairness algorithm is originally proposed for MANETs with the assumptions that malicious nodes always shows malicious behavior and node-to-node communication is always 100% successful without any link failures. But in real world deployment of community WMNs, we could not make the same assumptions because malicious user could show a probabilistic malicious behavior and node-to-node communication could always be affected by link failure. Hence MPIFA introduces some modifications to the original PIFA algorithm in order to achieve optimum fairness performance in Community WMN scenario.

**Probabilistic Malicious Behavior**

When malicious nodes show higher probability of malicious behavior, Original PIFA is able to remove 100% of malicious nodes within short period of time as seen in simulation results in Fig 5. But this effectiveness changes drastically with lower probability of malicious behavior when using Original PIFA where nodes do not drop packets each time they come across as assumed by the PIFA algorithm. Time taken to increase the NAM count of this particular malicious node over the blacklisting threshold goes up since it does not commit malicious acts often allowing it to disrupt the network operations for a longer period of time. From simulations we observed with malicious behavior probability of 30% it takes as much as 3 times more compared to the time of 100% malicious probability to remove them completely from the network.

This is much realistic scenario in Community WMN as some users tend to use the network unfairly; mostly during peak period where higher network utilization would costs more credits to these particular nodes and operating regular manner in other times. In a social network, we could expect more users to behave with partial malicious behavior rather than being completely unfair to the network. This behavior could results in lower End Delivery Ration (EDR) which in turn means lower throughput in the network.

This issue has been addressed by introducing modifications and Modified PIFA could detect the malicious nodes even with smaller malicious behavior probability much faster and remove them from the network. When malicious nodes show a probabilistic behavior, the NAM count increment is much slower for those nodes. But we could safely assume that malicious node do produce more inconsistence messages to CMS than a fair regular node. With this assumption, we could make malicious nodes NAM count run over the threshold much faster if we change the PIFA algorithm to increment NAM count faster for each inconsistency detected in a nodes' message(lesser number of inconsistence reports are produced due probabilistic behavior) by changing the NAM increment policy (5).

$$m_{a,b} = m_{a,b} + X$$
$$m_{b,a} = m_{b,a} + X \quad (7)$$

where, X is the NAM increment limit(>1) and increasing this limit would improve the malicious nodes detection time.

This modification (7) also could affect the non malicious nodes, specially neighbors of the malicious nodes because of the false allegations done by CMS. False allegation has to be normalized using NAM decrease policy (6) where NAM of previously alleged nodes with respect to currently alleged nodes should be reduced by a larger amount.

$$m_{i,a} = m_{i,a} / Y$$
$$m_{i,b} = m_{i,b} / Y, \text{ for all 'i' except (a,b)} \quad (8)$$

Y is the NAM reduction factor which we have to manipulate in order to avoid fair nodes being penalized (8). It is important to find the balance between NAM increment and decrement for an allegation in order to remove the malicious nodes from the network without effecting fair neighbors.

**Link Failures**

Original PIFA algorithm has another inherent assumption of 100% successful node-to-node transmissions without any link failures. But when adopting PIFA to a Community WMN, this assumption becomes obsolete as in real world conditions, link failures do occur rather often depending on the outdoor conditions. Community WMN is deployed generally in metropolitan areas or rural areas to provide network access to users in neighborhoods. This means that node-to-node communications are not guaranteed to be 100% successful as links could be disrupted for numerous natural and non-natural reasons.

But Original PIFA faces a problem when link failures do occur because originally it could not differentiate between inconsistent reports produced by intentional packet drops by malicious nodes and link failure packet drops by non-malicious node. When a fair node forwards a packet, the node databases are updated to account for the forwarded packets even though packets are not received by the next hop neighbor due to a failed link. Since there are no acknowledgements exchanged in these communications, both nodes are unaware of link failure. CMS perceive this inconsistency as node manipulations and penalize both nodes. Consequently many non-malicious nods are charged as malicious and blacklisted from the network when higher degree of link failure presents in the network.

Even though we could solve this by implementing acknowledgement messages during each exchange, it is not convenient as it would require the manipulation of physical layer of the network. Since our objective is to achieve optimum performance without disturbing lower protocol layers, we could achieve a considerable performance improvement by simply manipulating CMS updating frequency of MPIFA. When link failures are present in the network, both malicious nodes are well as link failures produce inconsistence reports incrementing NAM of the nodes involved. Our objective is to keep the malicious nodes NAM count constant while trying to get the NAM count of the non-malicious nodes down faster so that those nodes are not penalized due to link failures. It is important to keep in mind that non-malicious nodes in this case accumulate NAM, both due to inconsistency originated against malicious nodes and due to link failure. Hence it is important to reduce the NAM of non-malicious nodes faster before they get blacklisted. Best way to do this is to utilize the NAM decrease policy that has already been implemented in the system and optimized for the nodes probabilistic behavior.

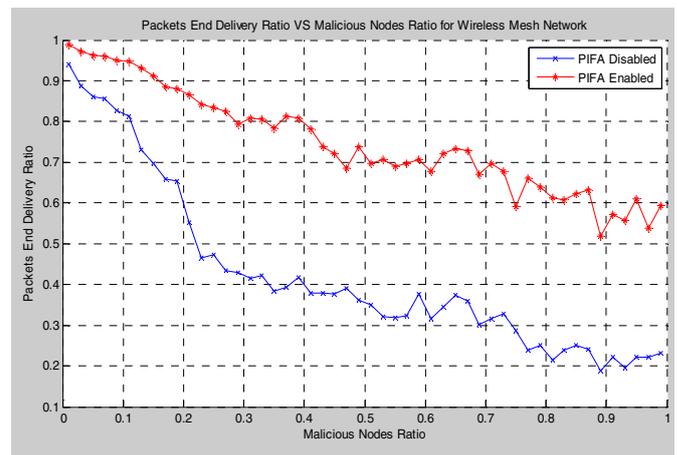

Fig 3: EDR vs MNR for WMN

By updating CMS in much faster frequency gives the opportunity for NAM decrease policy (8) to be activated in CMS more frequently thus reducing the NAM count of the falsely alleged non-malicious nodes to reduce in number of false detections significantly.

## V. SIMULATION METHODOLOGY

In order to evaluate the performance of the algorithm in Community WMN, simulations were carried out using MATLAB programming platform. This simulation omits physical layer characteristics of the network and use simpler routing methods since Original or Modified PIFA does not depend on the underlying routing protocols. Higher level language like MATLAB gives much simpler approach as oppose to lower level languages such as NS-2 when simulating this concept since we are carrying out higher layer (Application layer) protocol simulation. General WMN simulator was modeled initially and then Original PIFA was implemented on top of basic network in order to utilize the simulation effectively. Then analysis was carried out to determine whether the concept of PIFA effectively work in WMN. After analyzing the possibility of using PIFA in Community WMN, Modified PIFA was implemented for a comparison with Original PIFA.

During the simulation, a random node distribution of 50 nodes in 1000m x 1000m area with each node having a 300m wireless connection radius was used.

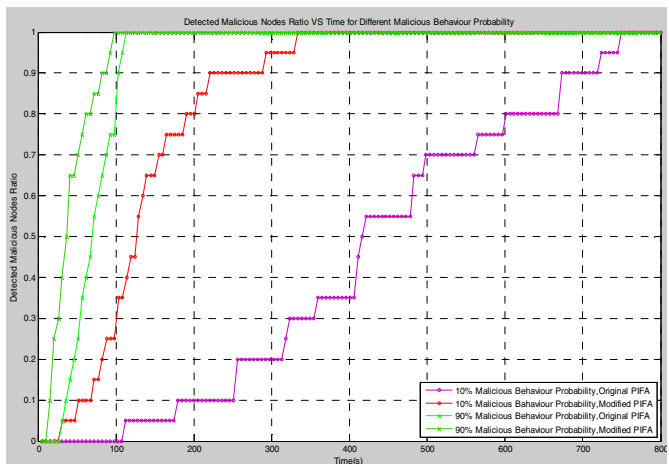

Fig 4: Performance comparison of Detected MNR with time taken for detection

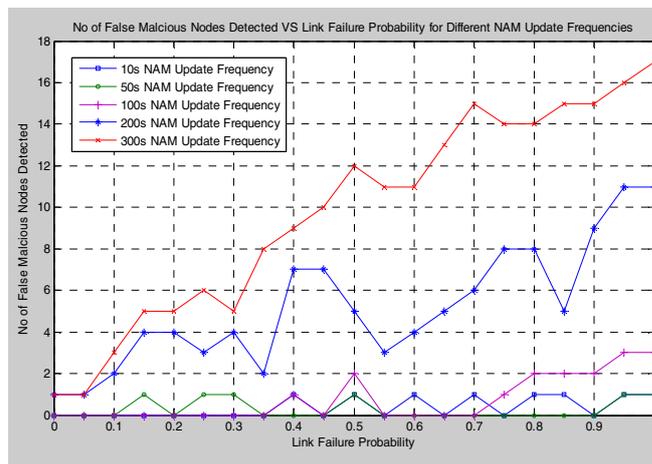

Fig 5 : False node detection with link failure for different CMS update frequency

## VI. PERFORMANCE EVALUATION of PIFA in WMN

Two key terms are used during the performance evaluation. Packets End Delivery Ratio (EDR) is the ratio of total packets received by the destination to total packets transmitted by non-malicious nodes. Malicious Nodes Ratio (MNR) is the ratio of total number of malicious nodes to total nodes in the mesh network. In Fig 3 we could see the EDR for WMN nodes with and without PIFA implemented. When MNR increases, EDR drops significantly in WMN without PIFA because selfish nodes within the network drop other users' packets. More malicious nodes mean more packets drops delivering lesser and lesser packets to the destination. But when PIFA is implemented, we could observe higher EDR even with large number of malicious nodes in the Community WMN. Algorithm is able to keep EDR above 50% all the time by removing these malicious nodes from network quickly and allowing routing algorithm to ignore these nodes permanently in the routing decision. This results shows that PIFA algorithm could be effectively used in Community WMN even though it was originally designed for MANET.

## VII. OPTIMIZED PERFORMANCE

Second part of the simulation was carried out to compare performance between Original PIFA and Modified PIFA. First simulation was carried out to analyze and compare the malicious node detection time in the network with malicious nodes behavior probability.

As in Figure 4 our simulation shows that Modified PIFA algorithm can identify and eliminate malicious nodes much faster than Original PIFA algorithm. WMN with 50 nodes gives optimum results when NAM increment factor X is between 2 to 3 and decrement factor Y around 2.5. We could observe that with these optimizations even with 10% malicious behavior, nodes could be removed from the network substantially faster compared with original PIFA algorithm. When the malicious nodes are removed from the Community WMN faster, Network EDR performance could be enhanced significantly.

When link failures exist in the network, algorithm should be able to identify malicious behavior from link failures effectively. By using the techniques introduced in Modified PIFA we could increase the performance of the algorithm by reducing the number of false detections significantly. By changing CMS update frequency appropriately, false node detections are reduced to a very minute number.

The simulation results in Fig 5 are achieved for a WMN with 50 nodes where 50% of them are malicious. According to the observations, by using CMS update frequency changing technique, we could achieve very acceptable low false detection rate even with a higher link failure rate. But we have to take in to account that higher frequency of CMS update means more overhead in the network, utilizing more bandwidth. Thus optimum NAM update frequency has to be utilized which gives the best possible tradeoff between lower overhead in the network and node-malicious nodes false detections.

## VIII. CONCLUSION AND FUTURE WORK

Implementation of a method to encourage users to be fair in network operations and identify malicious nodes would prove to be crucial to achieve maximum throughput and QoS of a community WMN. We proposed to use Modified Protocol Independent Fairness Algorithm in Community WMN to encourage nodes to be fair in packet forwarding and identify and remove malicious nodes in the network. We could observe from simulations that use of MPIFA in WMN effectively remove malicious nodes while encouraging every node to actively participate in packet forwarding. By removing malicious nodes in WMN, throughput and QoS could be dramatically improved in the network. By introducing MPIFA algorithm to Community WMN we could achieve far superior performance compared to Original PIFA in the midst of non-to-node link failures and probabilistic behavior of malicious nodes which are common phenomenon of Community WMN. Since the algorithm could be adapted in to existing routing protocols and physical layer protocols, implementation of proposed solution in real world operational networks would be straightforward.

For future work we propose methods to optimize the overhead of the messages to achieve best network utilization and incorporate characteristics of decentralized management in to the network to reduce delay of the network.

REFERENCE

[1] B.S. Manoj and Ramesh R. Rao, ''wireless mesh networks: issues and solutions'' in WIRELESS MESH NETWORKING : Architectures, Protocols and Standards, pp 3-48. 13 Dec 2006.

[2] Zhong, S., Chen, J. and Yang, Y.R., "Sprite: a simple, cheat-proof, credit-based system for mobile ad-hoc networks," INFOCOM 2003. Twenty-Second Annual Joint Conference of the IEEE Computer and Communications Societies. IEEE , vol.3, no., pp. 1987-1997 vol.3, 30 March-3 April 2003.

[3] Mogre, P.S., Graffi, K., Hollick, M. and Steinmetz, R., "AntSec, WatchAnt, and AntRep: Innovative Security Mechanisms for Wireless Mesh Networks," Local Computer Networks, 2007. LCN 2007. 32nd IEEE Conference on , vol., no., pp.539-547, 15-18 Oct. 2007.

[4] Buttyan, L. and Hubaux, J.-P., "Enforcing service availability in mobile ad-hoc WANs," Mobile and Ad Hoc Networking and Computing, 2000. MobiHOC. 2000 First Annual Workshop on , vol., no., pp.87-96, 2000.

[5] K. Lai, M. Baker, S. Marti and T. Giuli, "Mitigating routing Misbehavior in mobile Ad hoc networks," in Proceedings of MOBICOM 2000, pp. 255–265, 2000.

[6] Younghwan Yoo, Sanghyun Ahn and Agrawal, D.P., "A credit-payment scheme for packet forwarding fairness in mobile ad hoc networks," Communications, 2005. ICC 2005. 2005 IEEE International Conference on , vol.5, no., pp. 3005-3009 Vol. 5, 16-20 May 2005.